\documentclass{article}

\usepackage{arxiv}

\usepackage[utf8]{inputenc} 
\usepackage[T1]{fontenc}    
\usepackage{hyperref}       
\usepackage{url}            
\usepackage{booktabs}       
\usepackage{amsfonts}       
\usepackage{nicefrac}       
\usepackage{microtype}      
\usepackage{lipsum}
\usepackage{graphicx}
\graphicspath{ {./images/} }
\usepackage{amssymb}
\usepackage{amsmath}
\usepackage{hyperref}
\usepackage{xcolor}

\newcommand{\exv}[1]{\left\langle #1 \right\rangle}

\title{Water-methanol mixture confined in a graphene slit-pore}

\author{
 Roger Bellido-Peralta \\
  National Graphene Institute\\
  Univeristy of Manchester\\
  M13 9PL, Manchester \\
  \\
  Secció de Física Estadística i Interdisciplinària - Departament de Física de la Matèria Condensada\\
  Universitat de Barcelona\\
  08028, Barcelona \\
  \texttt{roger.bellido@postgrad.manchester.ac.uk} \\
   \And
 Fabio Leoni \\
  Department of Physics\\
  Sapienza University of Rome\\
  00185, Rome\\
  \texttt{fabio.leoni@uniroma1.it} \\
   \And
 Carles Calero \\
  Secció de Física Estadística i Interdisciplinària - Departament de Física de la Matèria Condensada\\
  Universitat de Barcelona\\
  08028, Barcelona \\
  \\
  Institut de Nanociència i Nanotecnologia\\
  Universitat de Barcelona\\
  08028, Barcelona \\
  \texttt{carles.calero@ub.edu} \\
  \And
 Giancarlo Franzese \\
  Secció de Física Estadística i Interdisciplinària - Departament de Física de la Matèria Condensada\\
  Universitat de Barcelona\\
  08028, Barcelona \\
  \\
  Institut de Nanociència i Nanotecnologia\\
  Universitat de Barcelona\\
  08028, Barcelona \\
  \texttt{gfranzese@ub.edu} \\
}

\begin{document}
\maketitle
\begin{abstract}
Efficient and sustainable techniques for separating water-methanol mixtures are in high demand in the industry. Recent studies have revealed that membranes and 2D materials could achieve such separation. In our research, we explore the impact of a nanoconfining graphene slit-pore on the dynamics and structure of water-methanol mixtures. By Molecular Dynamics simulations of a coarse-grained model for water mixtures containing up to 25\% methanol, we show that, for appropriate pore sizes, water tends to occupy the center of the pore. In contrast, methanol's apolar moiety accumulates near the hydrophobic walls. Additionally, modifying the pore's width leads to a non-monotonic change in the diffusivity of each component. However, water always diffuses faster than methanol, implying that it should be possible to identify an optimal configuration for water-methanol separation based on physical mechanisms. Our calculations indicate that one of the more effective pore sizes, 12.5\AA, is also mechanically stable, minimizing the energy cost of a possible filtering membrane. 
\end{abstract}


\section{Introduction} \label{intro}

Mixtures of water and methanol are widely utilized in various processes in the chemical  \cite{Miganakallu:2020aa}, food, and pharmaceutical industries  \cite{Shu2020}, as well as in biofuel production \cite{Masoumi:2021aa}. However, traditional methods for separating these mixtures, such as distillation, can be inefficient and expensive. As a result, researchers have turned their attention to membrane filtration technology as a potential solution for more cost-effective and efficient separation \cite{Mohammad2015}.

Graphene and its derivative, e.g., graphene oxide, are gaining significant interest due to their exceptional properties, including high flexibility, stiffness, and thermal conductivity \cite{Papageorgiou2017}. These properties make it a promising material for filtration technology. One crucial aspect of membrane filtration is the hydrophobicity or hydrophilicity of the material \cite{Kopel:2019aa}. However, applying simple processes such as chemical reduction \cite{Chua2014}, or thermal annealing \cite{Kumar2013} can alter hydrophilic materials to become more hydrophobic while still maintaining their excellent properties, ensuring high-quality results.

Significant progress has been made in the investigation of carbon-based materials. Experiments show that it is possible to adjust their semi-per\-mea\-bi\-li\-ty to various elements and molecules \cite{Nair2012} or functionalize them to improve their selectivities \cite{Chen2020, Wang2022, CastroMunoz2019}. Despite their potential applications, these materials have limitations. For example, graphene oxide swells when exposed to water \cite{Huang_2021}. Furthermore, scalability may only sometimes be feasible, limiting their use in industries, and some procedures required to obtain beneficial properties can be costly and time-consuming \cite{Smith2019, Jang2020}.
 
Molecular Dynamics (MD) simulations are a promising solution to overcome these limitations. They offer valuable insight into separating binary or multiple component mixtures, gas mixtures, and even water purification from wastewater at the mesoscale. Several studies, such as \cite{Gravelle2016, Yoo2017, Mangindaan2018, Zuo2021}, for example, have shown the potential of MD simulations in this regard.

The joint efforts of many research groups in the field of MD simulations of confined water, e.g., \cite{Engstler:2019aa, Corti:2021uy, Mendonca:2023aa} to mention a few, as well as our previous MD studies, have demonstrated that confinement at the nano-level can alter the behavior of a mono-component fluid compared to its behavior in the bulk \cite{LF2014, Leoni:2016aa, calero2020, Leoni:2021aa}. Building on these results, here we consider a binary mixture to model water-methanol, nanoconfined in a graphene slit pore and specifically examine the effects of nano-pore size on the acceptance capacity, density profile, diffusion coefficient, and hydration pressure.
    
\section{Models and Method} \label{M&S}

\begin{figure}
\centering
\includegraphics[width=1\columnwidth]{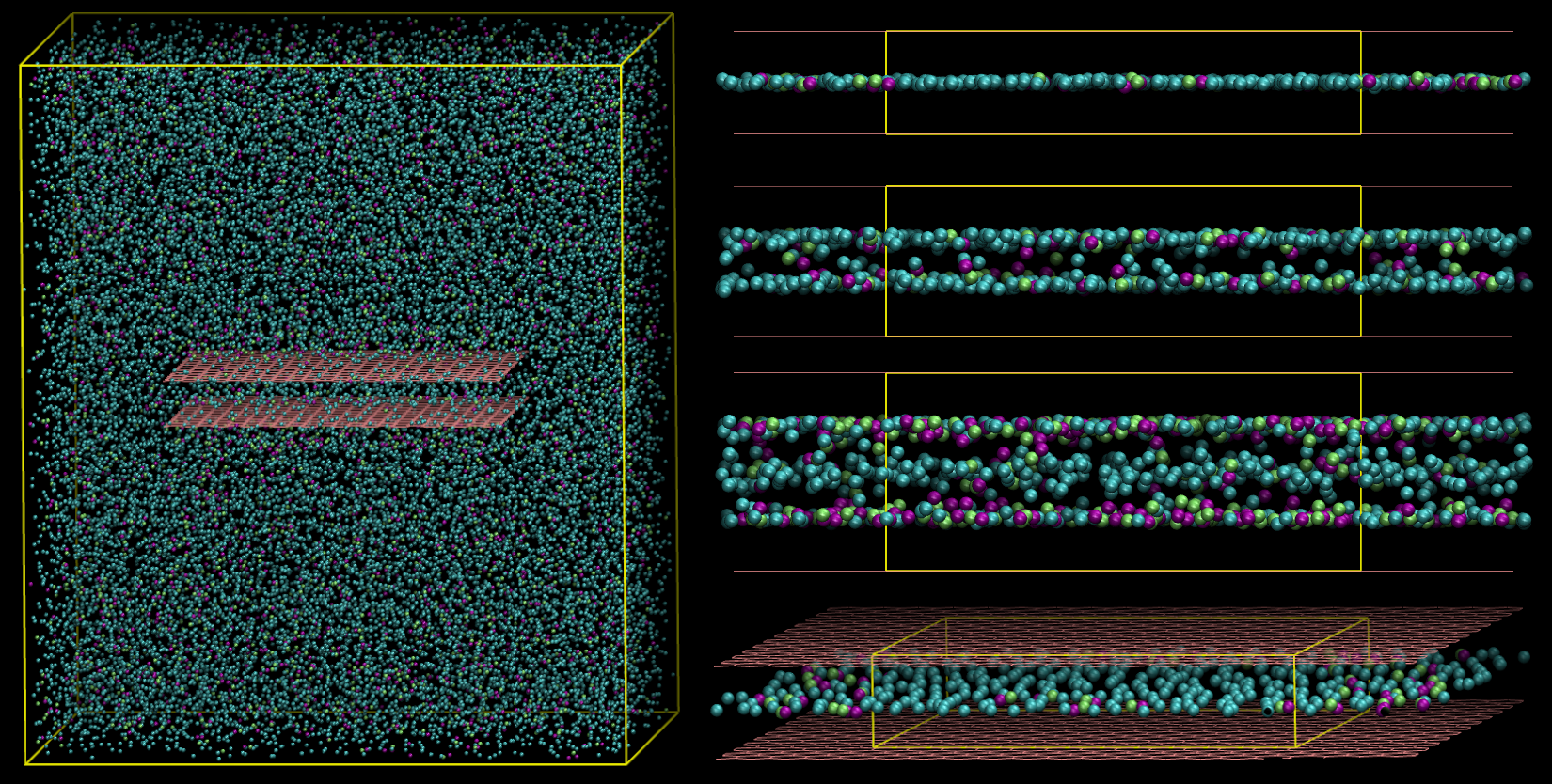}
\caption{{\bf Simulation snapshots for a 90\%-10\% CSW-methanol mixture}. Left: Simulation box (in yellow) with the graphene slit-pore in its center (pink). Right, from top to bottom: side view of a mixture monolayer in the slit pore with $\delta=6.5\,$\AA; a bilayer for $\delta=9.5\,$\AA; a three-layer for $\delta=12.5\,$\AA; a different view of the monolayer to emphasize the subvolume $V_s$ (in yellow) in which we compute the observables. Beads are water-like CSW particles (blue), methanol methyl groups (purple), and methanol hydroxyl groups (green). Pink lines represent the graphene lattice.}
\label{fig:sim}
\end{figure}

Our study examines a liquid mixture embedding a slit pore consisting of two parallel graphene walls of sizes 
$l_{x,gr}=49$\AA, 
$l_{y,gr}=51$\AA,
and separated by a fixed distance 6.5\AA\ $\leq \delta\leq$ 17\AA. 
The walls are composed of fixed carbon atoms arranged in a graphene structure and interacting with the liquid mixture through a 12-6 Lennard Jones (LJ) potential that aligns with the CHARMM27 force field \cite{Leoni:2021aa}.
To minimize edge effects of the walls, we calculate the observables for the confined mixture within a reduced region of the slit-pore, i.e., a central subvolume $V_{s}=L_{x,s}\, L_{y,s}\,\delta$, where 
$L_{x,s}=L_{y,s}=30$\AA\   (Fig. \ref{fig:sim}). 

We use coarse-grain models to represent the two mixture components, water and methanol. This approach reduces the computational time required for the calculations while reproducing consistent thermodynamic results \cite{BLCF_2023}. As per the works of Hus et al. (2014), Munao et al. (2015), and Marques et al. (2020)  \cite{Hus:2014aa, Munao:2015vo, Marques:2020aa}, we represent the methanol using two rigidly bonded beads - one for the apolar methyl group (CH$_3$) and the other for the polar hydroxyl group (OH). A 24-6  LJ potential models the interaction energy of two apolar groups while a Continuous Shouldered Well (CSW) potential \cite{Fr07a, Vilaseca2010} that of two polar moieties. The CSW model shows some of the structural, dynamical, and thermodynamic anomalies typical of hydrogen-bonding liquids \cite{oliveira2008, Vilaseca2010, Vilaseca2011}. 

To facilitate the blending of methanol with water, which forms hydrogen bonds with it, for consistency, we also model water via the CSW. However, due to its isotropy, the CSW cannot fully replicate the entropy behavior associated with the hydrogen-bond network \cite{oliveira2008, Vilaseca2010, Vilaseca2011}. For instance, in a hydrophobic graphene pore, water experiences a free energy minimum corresponding to a confined double layer dominated by potential energy and a secondary minimum for monolayers driven by entropy \cite{calero2020}. In contrast, in the CSW, both the monolayer and the bilayer minima are dominated by entropy \cite{Leoni:2021aa}. Nonetheless, despite these limitations, the combined representation of water and methanol effectively captures the primary characteristics of the mixture \cite{BLCF_2023}. We will elaborate on this further in the following discussion. The definitions and parameters of the potentials are given in Ref.~\cite{BLCF_2023}.

To compute the density profile along the direction $z$, perpendicular to the walls, with the origin in the center of the pore,
we divide $\delta$ into $2000$ bins of equal size $dz$ and calculate
\begin{equation}
    \rho_\alpha(z)\equiv \frac{
    \exv{N_\alpha(z)}
    }
    {L_{x,s}\, L_{y,s}\,dz},
\end{equation}
where $\exv{\cdot}_\alpha$ is the temporal average at the equilibrium, and $\exv{N_\alpha(z)}$ is the average number of molecules of species $\alpha$ within a slab with thickness $dz$ and size $L_{x,s}\, L_{y,s}$ centered in $z$ along the pore. 
The possible species for $\alpha$ are CSW, methanol, or both.

We estimate the mean square displacement (MSD) parallel to the walls--for each species $\alpha$--as
\begin{equation}
\begin{split}
    \Delta_\alpha^2(t)\equiv \exv{(x(t+t_0)-x(t_0))^2+(y(t+t_0)-y(t_0))^2}_\alpha = \\ 
                        =  N_\alpha(t_0)^{-1}\sum_{i=1}^{N_\alpha} [(x_i(t+t_0)-x_i(t_0))^2+(y_i(t+t_0)-y_i(t_0))^2],
\end{split}
\end{equation}
where the averages are calculated over the center of mass of each molecule of the species $\alpha$, and $N_\alpha(t_0)$ is the number of molecules of that species in the volume $V_s$ at the time $t_0$.
  
From the large-time (diffusive) regime limit of the MSD, we calculate the parallel diffusion coefficient
\begin{equation}
    D_\|(\alpha) \equiv 
    \left[\frac{\Delta_\alpha^2(t_l)}{4t_l}\right]_{t_l> t_D}.
\end{equation}
For all our simulations, we check that $t_D=2$ ps is enough to explore the diffusive regime (Fig.~\ref{fig:S1}).
Factor 4 accounts for the two dimensions of the planar diffusion, and $t_l$ is the lag time, i.e., the time the molecules spend in the subvolume. Because we are interested in $D_\|$ within the confined subvolume $V_{s}$, we use for the calculation only the particles that remain within $V_{s}$ along the whole simulation.

Furthermore, we compute the acceptance capacity of the pore, defined as the ratio between the number of molecules of the species $\alpha$ inside the subvolume $V_{s}$ and its area
\begin{equation}
    \sigma(\alpha)\equiv \frac{\exv{N_\alpha}}{A_s},
    \label{sigma}
\end{equation}
where $A_s\equiv L_{x,s}\, L_{y,s}$, and the average number density inside the subvolume 
\begin{equation}
    \rho(\alpha) \equiv \frac{\exv{N_\alpha}}{V_s}.
    \label{rho}
\end{equation}

Finally, we compute the hydration pressure as the difference between the normal pressure $P_\perp$ exerted by the confined particles over the slit-pore walls and the bulk pressure $P_{\rm bulk}$
\cite{LF2014,Todd_1995,Varnik_2000}
\begin{equation}
    P_{\rm hydr}\equiv P_{\perp}-P_{\rm bulk},
    \label{P_hydr}
\end{equation}
where
\begin{equation}
    P_{\perp} \equiv \frac{1}{2A}\exv{\sum_i (F_{i,\,b}-F_{i,\,t})},
\end{equation}
 $F_{i,\,b}$ and $F_{i,\,t}$ are the forces exerted by the particle $i$ on the bottom and the top graphene walls, respectively, and $A$ is the area of the walls.

We perform simulations in the canonical ensemble with a fixed total number $N$ = 25,000 of particles composing the mixture, in a total volume $V=84 \times 84\times 98$~\AA$^3$ with periodic boundary conditions, at~fixed temperature $T=100$~K controlled by the Nos\'e--Hoover thermostat, as implemented in the LAMMPS software~\cite{LAMMPS}. The temperature is chosen to allow us to compare the present results with the previous data for the pure CSW case \cite{Leoni:2021aa}. 
Using the Leap--Frog integration algorithm \cite{Birdsall_2018}, a standard second-order and time-reversible integration method, with timestep $\delta t= 1$ fs,  we equilibrate the system for 1 ns ($10^6$ MD steps) starting from an initial configuration with two intertwined hexagonal lattices of CSW beads and methanol dimers. We gather 1000 datapoints along the simulation to calculate observables.
To accurately analyze the short and long times of the simulation, we use an uneven distribution of time steps to record the particle trajectories. Further details are presented in Ref. \cite{BLCF_2023}.

\section{Results and Discussion} \label{R&D}

\subsection{Acceptance capacity and density inside the pore.}

\subsubsection{The pure CSW case.}

\begin{figure} 
\centering
\includegraphics[width=1\columnwidth]{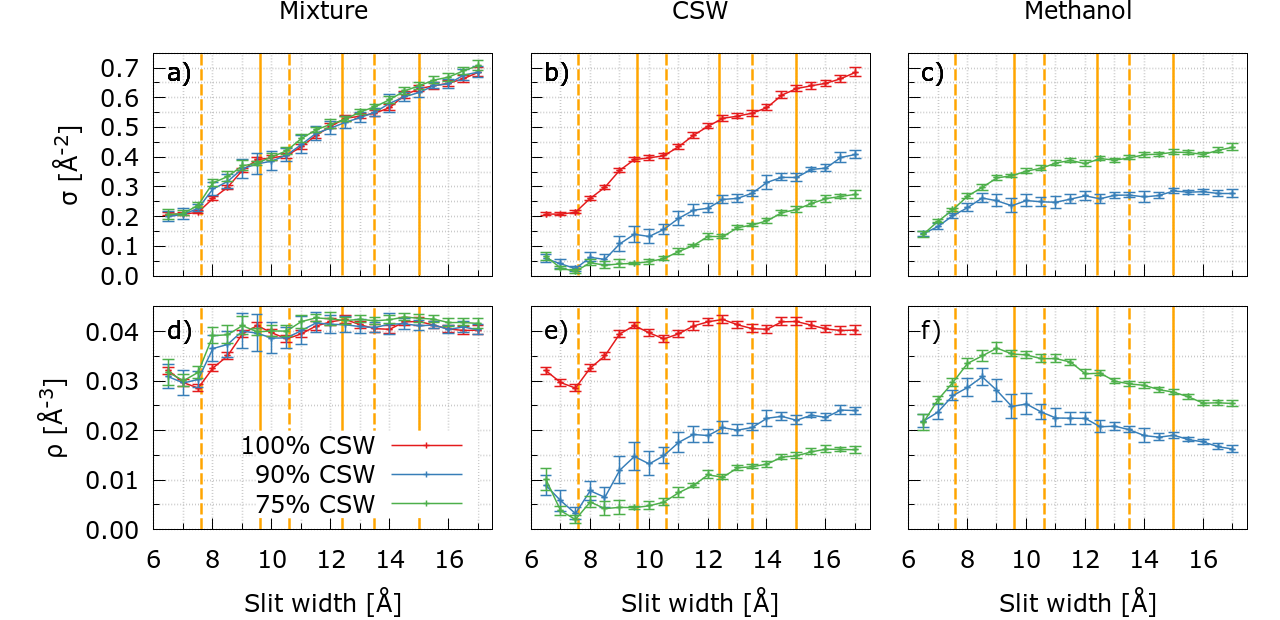}
\caption{{\bf  Acceptance capacity and fluid density in the slit-pore changing width.}
The acceptance capacity $\sigma$ (upper panels) and the fluid density $\rho$ inside the pore (lower panels) for the CSW-methanol mixture (panels a, d),
for compositions 100\%-0\% (red), 90\%-10\% (blue), and 75\%-25\% (green), 
have non-monotonic contributions for the CSW (panels b, e) and the methanol (c, f) components as a function of the slit-pore width $\delta$. Vertical solid (dashed) lines mark approximately local minima (maxima), within the present numerical resolution, of the pure CSW free energy \cite{Leoni:2021aa}.
}
\label{fig:acc_dens}
\end{figure}

To analyze the current findings, we first refer to the results obtained for the pure CSW under confinement \cite{Leoni:2021aa}. When $\delta$ is commensurable with an integer number of layers (solid vertical lines in Fig. \ref{fig:acc_dens} for the pure CSW case \cite{Leoni:2021aa}),
$\sigma$ remains constant regardless of minor changes in $\delta$.
These specific values of $\delta$ correspond to the lowest points in the pure CSW free energy and are mechanically stable, as the associated hydration pressure within the slit-pore vanishes \cite{Leoni:2021aa}. This is consistent with our current findings, which will be further discussed in section 3.4.
Although a slight decrease in $\delta$ near a stable point does not impact $\sigma$ since $N_\alpha$ remains constant, thanks to the softness of the CSW, it induces a maximum in $\rho$ because of the change in $V_s$ and in the number of layers. 

In contrast, if $\delta$ corresponds to a maximum in the CSW free energy (dashed vertical lines in Fig. \ref{fig:acc_dens} for the pure CSW case \cite{Leoni:2021aa}), the hydration pressure is zero only at a point of mechanical instability. As $\delta$ increases around this value, the pressure quickly transitions from negative (indicating effective attraction between walls) to positive (effective repulsion between walls) \cite{Leoni:2021aa}.
Therefore, a slight increase of $\delta$ near these mechanically unstable values causes a quick rise in both $\sigma$ and $\rho$ due to the resulting increase in $N_\alpha$. 

As the width of the slit-pore increases, the plateaus and extreme points of $\rho$ and $\sigma$ become less pronounced due to the reduced structure of the liquid inside the pore. This behavior continues until the density $\rho$ approaches a value similar to the overall density of the system.

\subsubsection{The mixture case.}

When both liquids are in the pore, we find that the mixture's acceptance capacity $\sigma$ and density $\rho$ do not change their qualitative behavior compared with the pure CSW case (Fig. \ref{fig:acc_dens} a, d).
However, within our numerical precision, both quantities increase slightly upon adding methanol.

The analysis of the behavior of each component reveals that the water-like liquid in the mixture is displaced from the inside of the pore compared to the pure case. The CSW acceptance capacity and density between the walls decrease significantly for all $\delta$ values (Fig. \ref{fig:acc_dens} b, e).
Furthermore, the local maxima and minima in $\sigma$ and $\rho$ are recovered for the same values of $\delta$ as in the pure CSW case, albeit with less clarity. 

We observe a more intriguing pattern for the methanol in the pore.  When the pore size is less than 9.0~\AA, the amount of methanol inside the pore significantly increases with $\delta$. However, above this pore-size threshold, $\rho$ reaches a maximum for both methanol concentrations, and the acceptance capacity levels off. The maximum in $\rho$ and the level in $\sigma$ increase for larger methanol concentration, overcoming the values of the CSW component and inverting their overall relative concentrations.

Finally, when $\delta$ is increased, the value of $\sigma$ for methanol stabilizes, but the value for the CSW continues to rise. As a result, the alcohol density inside the pore decreases gradually when the pore size exceeds 9.0~\AA (Fig. \ref{fig:acc_dens} c, f). These results indicate that the methanol, sequestered inside the slit pore from the bulk reservoir, has higher adsorption inside the structure than the CSW, as the density of methanol is larger (at almost every plate separation) than the water-like one despite representing a much lower percentage of the mixture (see Fig. \ref{fig:acc_dens} e, f).

\subsection{Density profile}

\begin{figure}
\centering
\includegraphics[width=0.8\columnwidth]{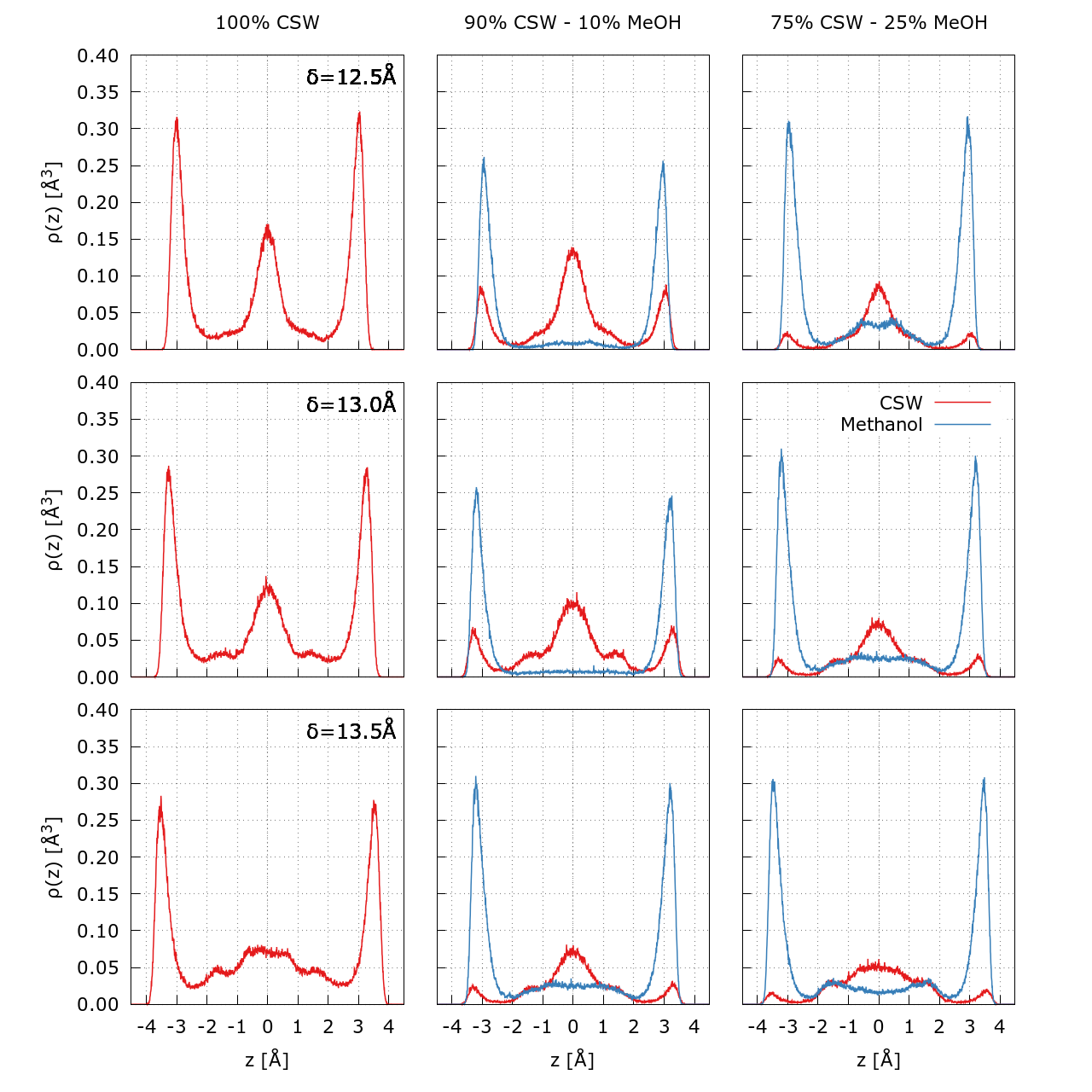}
\caption{{\bf The density profiles in the slit pore show methanol segregation.}
Density profiles $\rho(z)$ along the $z$ direction orthogonal to the pore walls for large slit pores with $\delta =12.5$ \AA\ (upper panels), 13.0 \AA\ (central row) and 13.5 \AA\ (lower panels) and CSW-methanol mixture compositions 100\%-0\% (left panels), 90\%-10\% (central column) and 75\%-25\% (right panels). The CSW water-like component (red) shows three layers. Instead, the methanol $\rho(z)$ (blue) shows only two layers, with significant depletion in the central region of the slit pore, especially for $\delta=12.5$ \AA\ and $\delta=13.0$ \AA\ at 90\%-10\% CSW-methanol composition.}
\label{fig:density}
\end{figure}

To better understand the behavior of $\sigma$ and $\rho$ inside the slit pore, we calculate the density profiles of each component at different values of $\delta$ and compositions (Figs. \ref{fig:density}, \ref{fig:S2} -- \ref{fig:S7}). At low methanol percentage (90\%-10\% mixture), we recover the layering observed in \cite{Leoni:2021aa} for pure CSW for all the pore widths. The methanol concentrates near the walls, thereby altering the density of the CSW in the same region and within the overall pore.

The larger methanol concentration near the wall results from its stronger interaction with the graphene atoms than the CSW. Indeed, in our model, the methanol dumbbells interact with the walls via twice the points of the CSW beads. This interaction creates a preferential attraction between methanol and the graphene sheets, similar to the hydrophobic collapse expected between carbon walls and methyl moieties. 

The methanol adsorption on the walls does not have a significant impact on the structure of the CSW in the center of the pore, where the methanol is almost absent in all those cases (with $\delta\geq$ 10 \AA) where there are at least three CSW layers (central panels in Figs. \ref{fig:density}, \ref{fig:S2} -- \ref{fig:S7}).
This is consistent with what is observed in atomistic simulations for water-methanol \cite{Prslja:2019aa} and water-ethanol mixtures  \cite{Mozaffari:2019aa} where the apolar CH$_3$ groups concentrate on the walls, leaving more space for water in the central layers. Therefore, although our model does not mimic the amphiphilic property of methanol because it lacks directional hydrogen bonds, it gives results consistent with detailed models.

Our analysis is also confirmed at the higher methanol concentration for the 75\%-25\%  mixture composition. Under these conditions, the concentration of alcohol in the pore increases, both in the central layer and near the walls, inducing a higher decrease in the CSW density, although the number of CSW layers does not change (right panels in Figs. \ref{fig:density}, \ref{fig:S2} -- \ref{fig:S7}). 

When the value of $\delta$ is less than 10 \AA, the CSW can only accommodate a maximum of two layers inside the pore along with the methanol (Figs. \ref{fig:S2} -- \ref{fig:S4}). This means that the presence of methanol in the pore significantly lowers the concentration of CSW, resulting in the highest density $\rho$ of methanol at $\delta\simeq$ 8.5 \AA, as seen in Fig.\ref{fig:acc_dens}. As $\delta$ further decreases, steric hindrance causes a reduction in the methanol $\sigma$ within the pore, lowering its $\rho$.

In particular, at  $\delta\simeq 7.5$ \AA, the density difference between the two components within the slit pore favors methanol the most (Fig.~\ref{fig:acc_dens}).  This corresponds to the pore size that exhibits the first maximum in the CSW free energy \cite{Leoni:2021aa}, which means that the CSW density inside the pore is also at its lowest in the pure case (Fig.~\ref{fig:acc_dens}). Thus, methanol can easily replace CSW in the pore under these circumstances. However, maintaining the pore size at this distance would require significant energy, as indicated by the maximum free energy.

On the other hand, when the size is $\delta=12.5$ \AA,  the separation between the two components is significant (Fig.~\ref{fig:density}), and the pore free energy is at a minimum 
(Fig.~\ref{fig:acc_dens}) \cite{Leoni:2021aa}. Therefore, the methanol segregation can be attained at this pore size in mechanically stable conditions with minimal energy cost.

Under these circumstances, a concentration gradient in the pore could prompt a lamellar flow, enabling the recovery of both components separately.
Moreover, even without a concentration gradient, the segregation effect could lead to a significant difference in the diffusion of the mixture's components. We will assess this prospect in the following section.

\subsection{Confined diffusion coefficient}

\begin{figure}
\centering
\includegraphics[width=1\columnwidth]{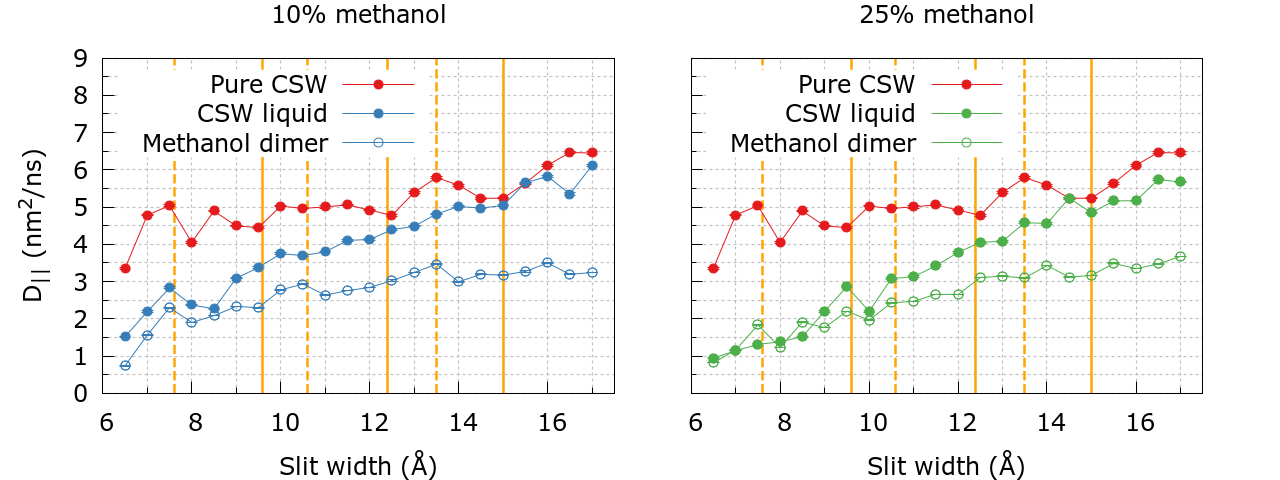}
\caption{{\bf Diffusion coefficient parallel to the slit pore walls for confined CSW-methanol mixtures.} The longitudinal $D_\|$ is non-monotonic with the pore width $\delta$. 
For pure CSW (red symbols and lines), it has maxima and minima near the free energy maxima and minima (marked by dashed and solid vertical thick lines, as in Fig.~\ref{fig:acc_dens}), respectively. 
For both 90\%-10\% (left panel) and 75\%-25\% (right panel) CSW-methanol mixtures, the water-like component (blue or green solid circles) is faster than the methanol (blue or green open circles), especially for large values of $\delta$. However, the partial mismatch of diffusion extrema with the pure case suggests, within our numerical errors, a complex dynamic interplay between the two components.
}
\label{fig:dmixt}
\end{figure}

We observe that the overall value of $D_\|$ decreases for each mixture component within the slit pore as $\delta$ is reduced (Fig. \ref{fig:dmixt}). 
This can be attributed to the fact that stronger confinement results in a reduced volume for diffusion. 
In the case of pure CSW \cite{Leoni:2021aa}, $D_\|$ displays oscillations with minima for those values of $\delta$ where the liquid reaches maximum density and maxima where the density is minimum (solid and dashed vertical thick lines in Fig.~\ref{fig:dmixt}, respectively). 

The extrema in $D_\|$ for the mixture's components are less evident and suggest an interplay between the two liquids.
For instance, CSW's $D_\|$ in the 90\%-10\% mixture decreases at $\delta\simeq 8.5$ \AA\ (Fig. \ref{fig:dmixt}) where methanol's $\rho$ is maximum (Fig.~\ref{fig:acc_dens}). This behavior could be a consequence of how the density of methanol impacts the diffusion of CSW in the mixture, specifically when their layers overlap ($\delta<10$ \AA, Fig.~\ref{fig:S3}, \ref{fig:S4}). This effect, however, is no longer evident for $\delta>$15 \AA, where the diffusion of CSW in the mixture becomes comparable to that of the pure case. 

We generally observe that the CSW is faster than the methanol within the pore. This could be due to the ability of the CSW soft-core potential to adapt quickly to pore widths that do not match with the components layers (dashed vertical thick lines for the pure CSW in Fig.~\ref{fig:dmixt}), as, for example, at $\delta\simeq 7.5$ \AA\ (Fig.~\ref{fig:S2}).
The partial overlaps of the soft cores of different molecules can lead to multiple maxima (and minima) in $D_\|$ at values of $\delta$ higher compared to the pure CSW case (Fig.~\ref{fig:dmixt}).

Also, as the methanol concentration increases in the mixture, the thermal diffusion of the water-like liquid becomes slower (Fig.\ref{fig:dmixt}). This phenomenon can be attributed to the steric effect of the voluminous methanol molecule, which restricts the availability of free volume for the diffusion of CSW. As a result, the value of CSW's $D_\|$ diminishes.

The extra-volume mechanism that facilitates thermal diffusion, due to the soft core, has a limited effect on methanol. This is because only one of its moieties interacts with the CSW potential. As a result, the methanol's maximum diffusion coefficient values are smaller than those of the water-like liquid. This observation is consistent with the fact that increasing the methanol concentration in the mixture or decreasing $\delta$ diminishes the soft-core facilitating effect, leading to a slowdown in the overall mixture (Fig. \ref{fig:dmixt}).

Furthermore, methanol experiences a more significant slowing down than the CSW component, which can be attributed to its stronger interaction with the walls. Methanol particles interact with the walls twice as much as CSW particles, leading to more effective hydrophobic attraction.

Therefore, a) the slowing down of methanol due to hydrophobic attraction and b) the facilitation effect of soft polar interaction combine to induce 1) CSW depletion from inside the pore for small $\delta$ values, 2) segregation between mixture components at higher $\delta$ values, and 3) generally faster longitudinal diffusion of the water-like liquid. These effects are promising prerequisites for efficient physical separation of the two components.

\subsection{Hydration Pressure}

\begin{figure}
\centering
\includegraphics[width=1\columnwidth]{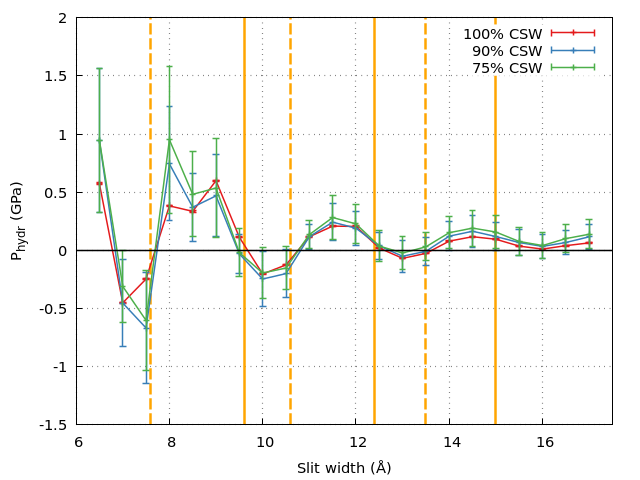}
\caption{{\bf The mixture's hydration pressure oscillates between positive and negative values depending on the slit-pore width.} Only a few values of $\delta$ correspond to mechanically stable (i.e., with $P_{\rm hydr}=0$ and $\partial P_{\rm hydr}/\partial \delta<0$, solid vertical thick lines) or mechanically unstable  (i.e., with $P_{\rm hydr}=0$ and $\partial P_{\rm hydr}/\partial \delta>0$, dashed vertical thick lines) equilibrium widths.
In between these values, external energy would be needed to keep $\delta$ fixed against the effective (mixture-mediated) wall-wall attraction ($P_{\rm hydr}<0$) or repulsion ($P_{\rm hydr}>0$). The range of variation of $P_{\rm hydr}$ is larger for the 
75\%-25\% CSW-methanol mixture (green symbols and lines) than for the
90\%-10\% mixture (blue), while the pure CSW case (red) is approximately intermediate. Vertical dashed and solid thick lines are as in Fig.~\ref{fig:acc_dens}.}
\label{fig:hydr}
\end{figure}

Next, we investigate the relationship between the hydration pressure $P_{\rm hydr}$ and $\delta$ to determine the most energy-efficient pore size for separation.  Indeed, $P_{\rm hydr}$ reflects the force required to maintain a specific distance $\delta$ between the walls of the slit pore. By integrating $P_{\rm hydr}$ along the $z$-axis, we can evaluate the amount of energy required to maintain $\delta$ fixed under the influence of external factors, such as an induced mixture flux.

Like water \cite{Engstler:2019aa, calero2020, Engstler:2018ab} and pure CSW liquid \cite{Leoni:2021aa}, the CSW-methanol mixture can either attract or repel the confining walls depending on the pore width $\delta$ (Fig.~\ref{fig:hydr}). If $P_{\rm hydr}>0$, the confined mixture's internal pressure on the pore walls is greater than the bulk pressure in Eq.(\ref{P_hydr}), and an external force is required to balance the internal pressure to keep $\delta$ constant and prevent the walls from moving far apart. Conversely, if $P_{\rm hydr}<0$, the bulk pressure is greater than the internal pressure, and an external force is necessary to avoid the collapse of the slit-pore.

As discussed for pure fluids, the force imbalance results from the system's changes in free energy. The stable equilibrium conditions ($P_{\rm hydr}=0$ and $\partial P_{\rm hydr}/\partial \delta<0$) correspond to the minima in free energy. In contrast, the unstable equilibrium conditions ($P_{\rm hydr}=0$ and $\partial P_{\rm hydr}/\partial \delta>0$) represent the maxima \cite{calero2020, Leoni:2021aa}.

An example that helps to understand this concept is when $P_{\rm hydr}=0$ under two conditions: at $\delta\approx7.5$ \AA\ and $\delta\approx9.5$ \AA. Our simulations show that two liquid layers form between these two values of $\delta$. If we decrease $\delta$ from $\delta\approx7.5$ \AA, the pressure becomes negative, and the slit-pore would collapse. Conversely, if we increase the width of the slit, $P_{\rm hydr}>0$ and the walls move apart. A similar but opposite situation occurs at $\delta\approx9.5$ \AA. If we decrease the width, the mixture applies $P_{\rm hydr}>0$ to restore the width to the equilibrium value. Conversely, if we increase $\delta$, $P_{\rm hydr}<0$ and the walls return to their original $\delta$ value. 

Here we find that the free-energy minima conditions for the mixtures are consistent, within the error bars, with those for pure CSW. Additionally, we observe that the maxima have a weak dependence on the percentage of methanol component (Fig.~\ref{fig:hydr}). Specifically, we find that the slit-pore is in stable equilibrium when $\delta\simeq 9.5$ \AA, 12.5 \AA\ and 16 \AA. These widths correspond to 1) a confined mixture bilayer with well-mixed CSW and methanol (Fig.~\ref{fig:S3} and \ref{fig:sim}), 2) three confined layers where CSW and methanol are segregated (Fig.~\ref{fig:density} and \ref{fig:sim}), and 3) four CSW layers with characteristics similar to the three-layer case.

Furthermore, in the case of three and four-layers of mixture, the water-like liquid has a thermal diffusion constant approximately 20\% to 50\% greater than methanol. However, in bilayer configurations, this difference diminishes, and it almost disappears when $\delta$ is decreased, with methanol concentration changes in the mixture having only a minor impact  (Fig.~\ref{fig:dmixt}).

Therefore, based on our findings, we identify two potential configurations that can be used for effective segregation processes. The first, at high energy cost, involves the mechanical instability surrounding $\delta \simeq 7.5$ \AA, where there is a significant difference in the amount of CSW water and methanol present within the pore. The second configuration is around $\delta\simeq 12.5$ \AA, which offers ideal segregation and diffusion conditions for mechanically separating the water-like liquid from methanol in an energy-efficient way.

\section{Summary and conclusions} \label{S&C}
The separation of water and methanol is essential in various industrial processes, such as extracting methanol for biofuels \cite{Ren_2000, Boysen_2004}. However, conventional techniques have their limitations regarding efficiency and costs \cite{Liang_2014}. Thus, it is crucial to seek alternative methods that are both technologically feasible and scientifically compelling. This is particularly relevant for achieving the UN Sustainable Development Goals (SDGs) of ``Clean Water and Sanitation" and ``Affordable and Clean Energy''.

Here, we aim to explore the potential of separating a water-methanol mixture by confining it in a graphene slit pore using Molecular Dynamics. We perform simulations of mixtures with varying compositions in slit pores of different widths. We use the CSW water-like liquid \cite{Fr07a} and the dumbbell methanol model \cite{Hus:2014aa} to coarse-grain the mixture to ensure simulation efficiency while retaining the correct thermodynamic behavior of the mixture \cite{BLCF_2023}. Both models have been previously tested in the literature \cite{oliveira2008, Munao:2015vo} and have successfully replicated mixture properties \cite{Marques:2020aa, Prslja:2019aa}.

Under appropriate conditions, we observe that the two liquids a) segregate and b) diffuse at different rates along the pore in a slit geometry. Due to its strong hydrophobic attraction with graphene, it is noteworthy that methanol tends to concentrate inside the slit pore. The interplay between 1) layering, 2) slowing down caused by the interaction of graphene and methanol, and 3) facilitation of diffusion due to soft polar interactions create optimal conditions for confining the mixture, leading to enhanced separation.

In particular, our findings suggest that separating water and methanol in a graphene pore is most effective when the pore width contains at least three mixture layers, equivalent to a width of approximately 12.5 \AA. At this pore size, three factors contribute to an energetically-favorable mixture separation.  I) The components segregate, concentrating the methanol near the walls and the water in the pore center. This setup allows efficient water extraction from the mixture through proper filtering geometry. II) The methanol is about 20\% slower than water in the concentration range of 10\% to 25\%, further aiding in the separation process. III) The pore remains mechanically stable, minimizing the energetic costs against external forces such as fluxes applied to the mixture. These results suggest the potential of exploring optimal pore parameters for the physical filtration and purification of water-alcohol mixtures through nanotechnology.

\vspace{6pt} 

\textbf{Funding} \\
This research was funded by MCIN/AEI/ 10.13039/ 501100011033 and ``ERDF A way of making Europe" grant number PGC2018-099277-B-C22 and PID2021-124297NB-C31.

\textbf{Acknowledgments} \\
G.F. acknowledges the Visitor Program of the Max Planck Institute for The Physics of Complex Systems for supporting a six-month visit that started on November 2022. \\
R. Bellido-Peralta would like to acknowledge the assistance given by Research IT and the use of the Computational Shared Facility at The University of Manchester.

 \bibliographystyle{ieeetr} 
 \bibliography{wat-met}

\clearpage

\begin{center}
{\bf\large{Supplementary Material}}
\end{center}

\clearpage

\setcounter{equation}{0}
\setcounter{figure}{0}
\setcounter{table}{0}
\setcounter{section}{0}
\makeatletter
\renewcommand{\theequation}{S\arabic{equation}}
\renewcommand{\thefigure}{S\arabic{figure}}
\renewcommand{\thetable}{S\arabic{table}}
\renewcommand{\thesection}{S\arabic{section}}

\begin{figure} 
\centering
\includegraphics[width=1\columnwidth]{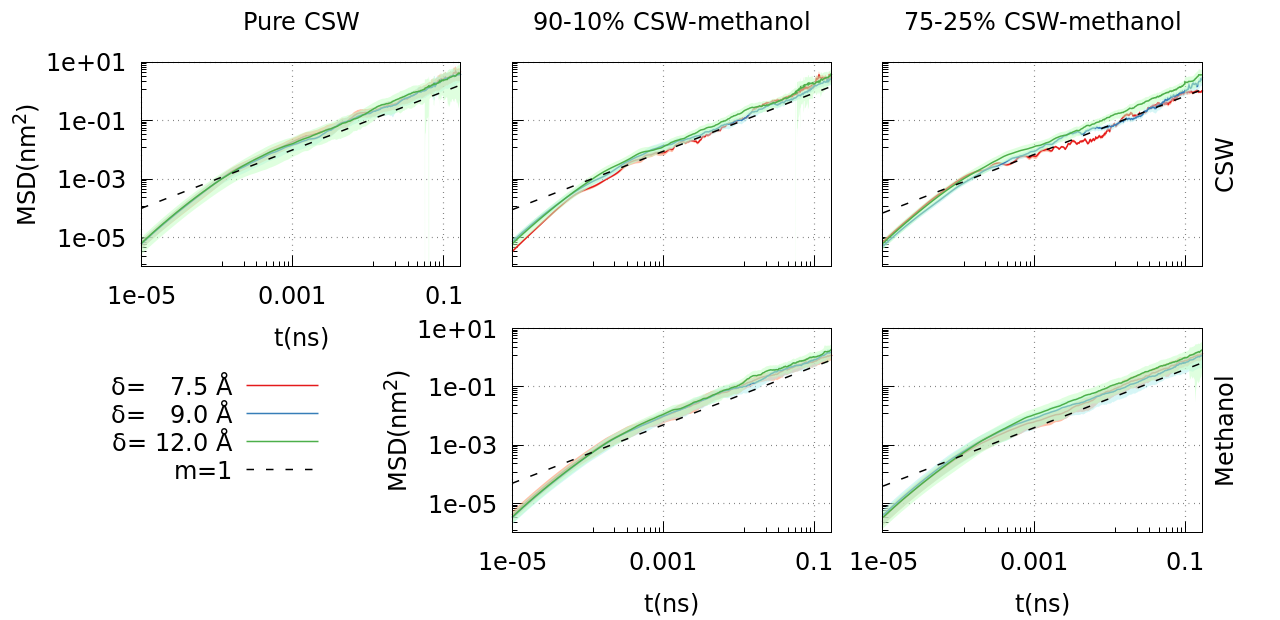}
\caption{{\bf MSD for confined CSW-methanol mixtures.}
For each slit pore width $\delta$ (7.5\AA, red lines; 9.0\AA, blue; 12.0\AA, green), both the CSW (upper panels) and the methanol (lower panels) reach the diffusive regime (power law with exponent 1, dashed lines) at any composition (pure, left panel;  90\%-10\%, central panels; 75\%-25\%, right panels) for $t>t_D=2$ ps.}
\label{fig:S1}
\end{figure}

\begin{figure} 
\centering
\includegraphics[width=1\columnwidth]{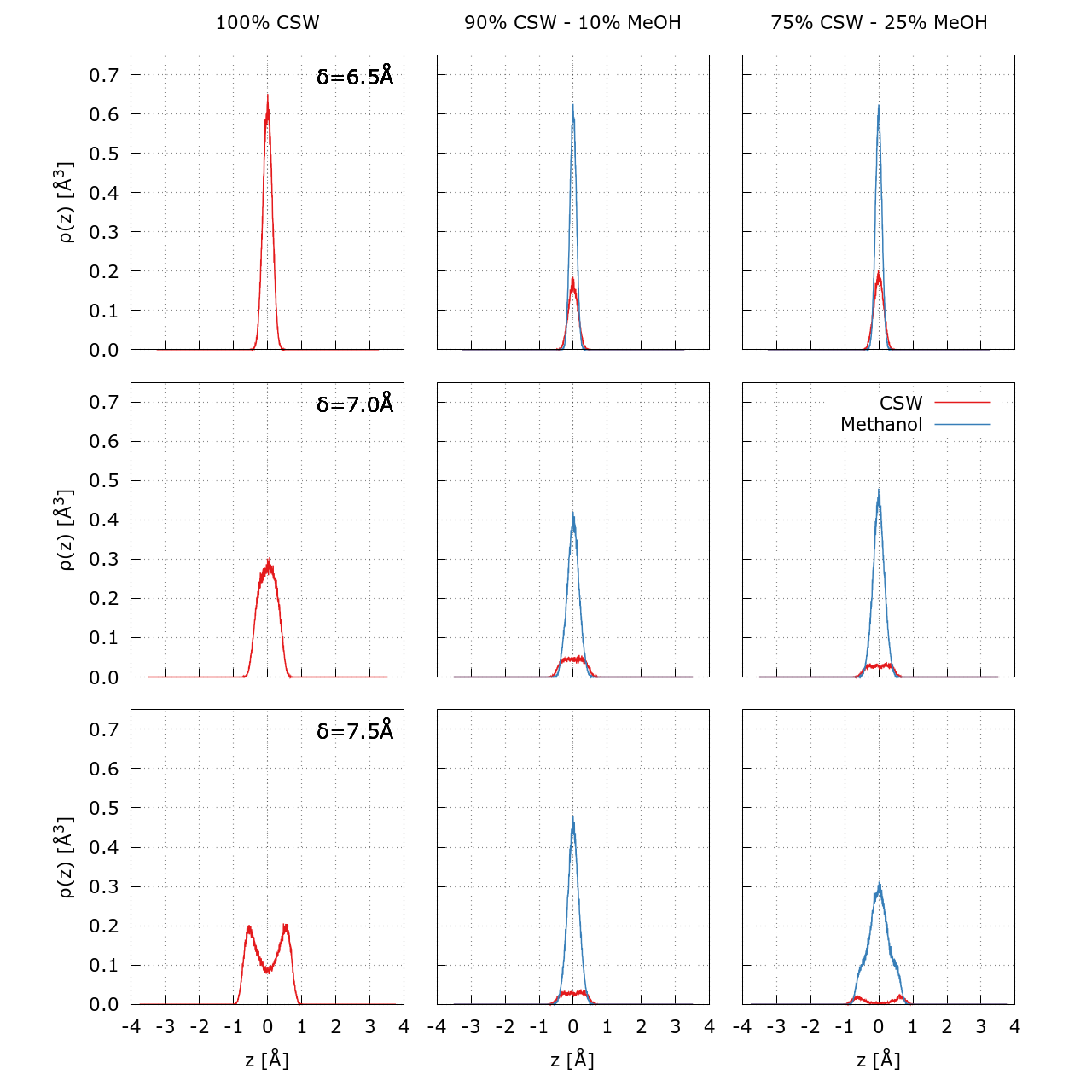}
\caption{{\bf Density profiles in the slit pore with at most one methanol layer.}
As in Fig.~\ref{fig:density}, but for 
$\delta =6.5$\AA\ (upper panels), 
7.0\AA\ (central panels), and 
7.5\AA\ (lower panels).
The CSW $\rho(z)$(red) increases from one to two at $\delta \simeq 7.5$\AA, while the methanol $\rho(z)$ (blue) always has one layer, although broadened for the larger width.} 
\label{fig:S2}
\end{figure}

\begin{figure} 
\centering
\includegraphics[width=1\columnwidth]{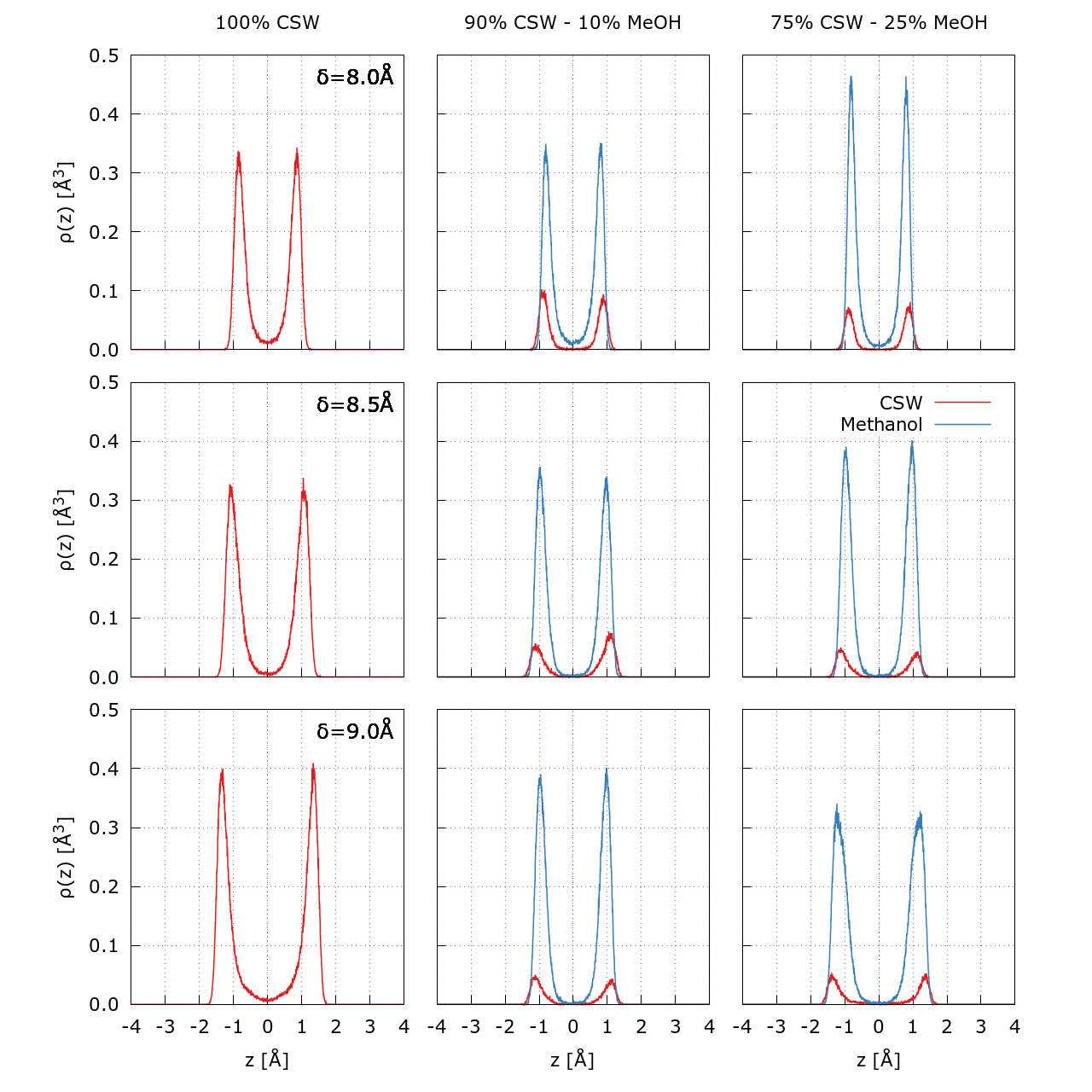}
\caption{{\bf Density profiles in the slit pore with two layers of CSW and two of methanol.}
As in Fig.~\ref{fig:S2}, but for 
$\delta =8.0$\AA\ (upper panels), 
8.5\AA\ (central panels), and 
9.0\AA\ (lower panels).} 
\label{fig:S3}
\end{figure}

\begin{figure} 
\centering
\includegraphics[width=1\columnwidth]{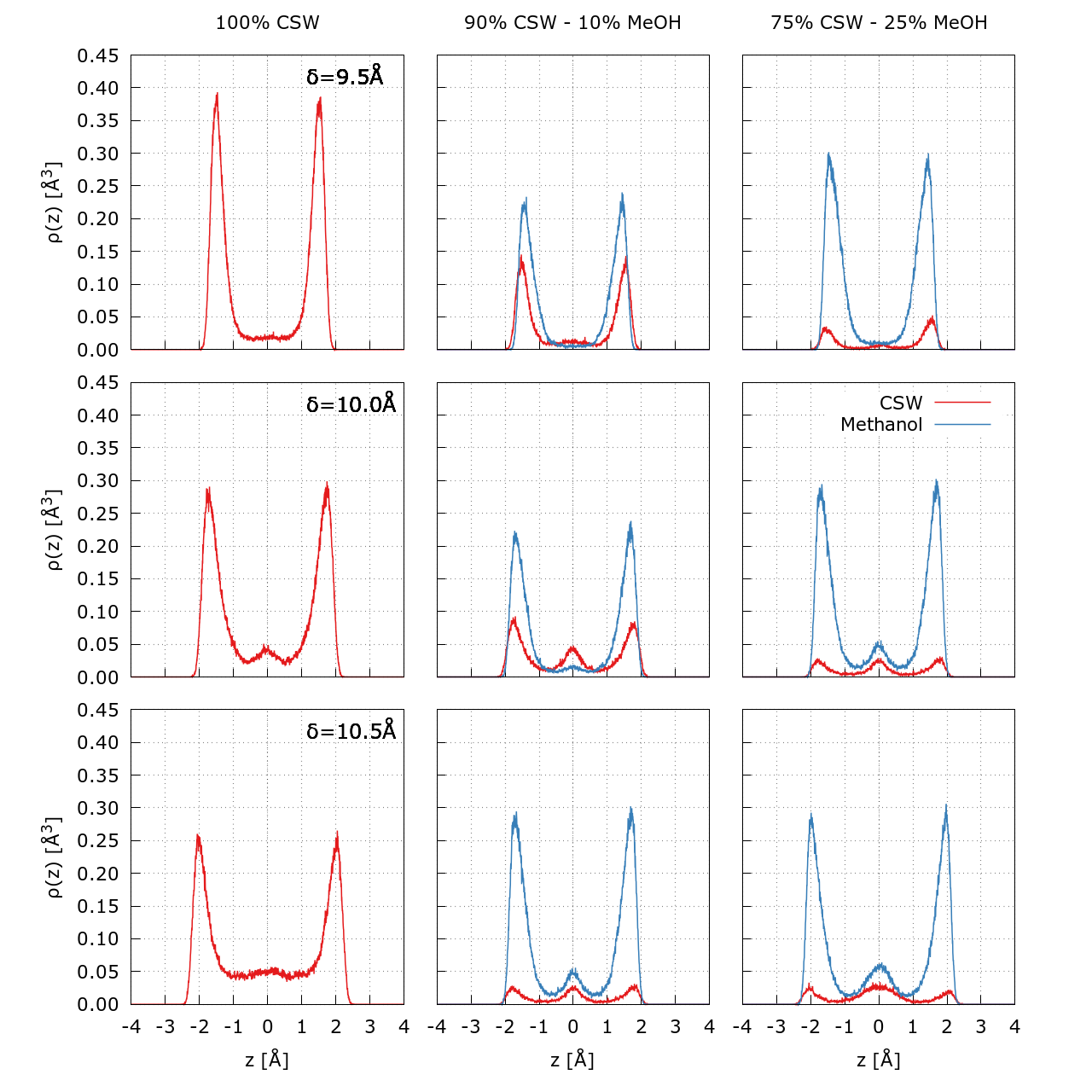}
\caption{{\bf Density profiles in the slit pore with an incipient third CSW layer in the center.}
As in Fig.~\ref{fig:S2}, but for 
$\delta =9.5$\AA\ (upper panels), 
10.0\AA\ (central panels), and 
10.5\AA\ (lower panels).} 
\label{fig:S4}
\end{figure}

\begin{figure} 
\centering
\includegraphics[width=1\columnwidth]{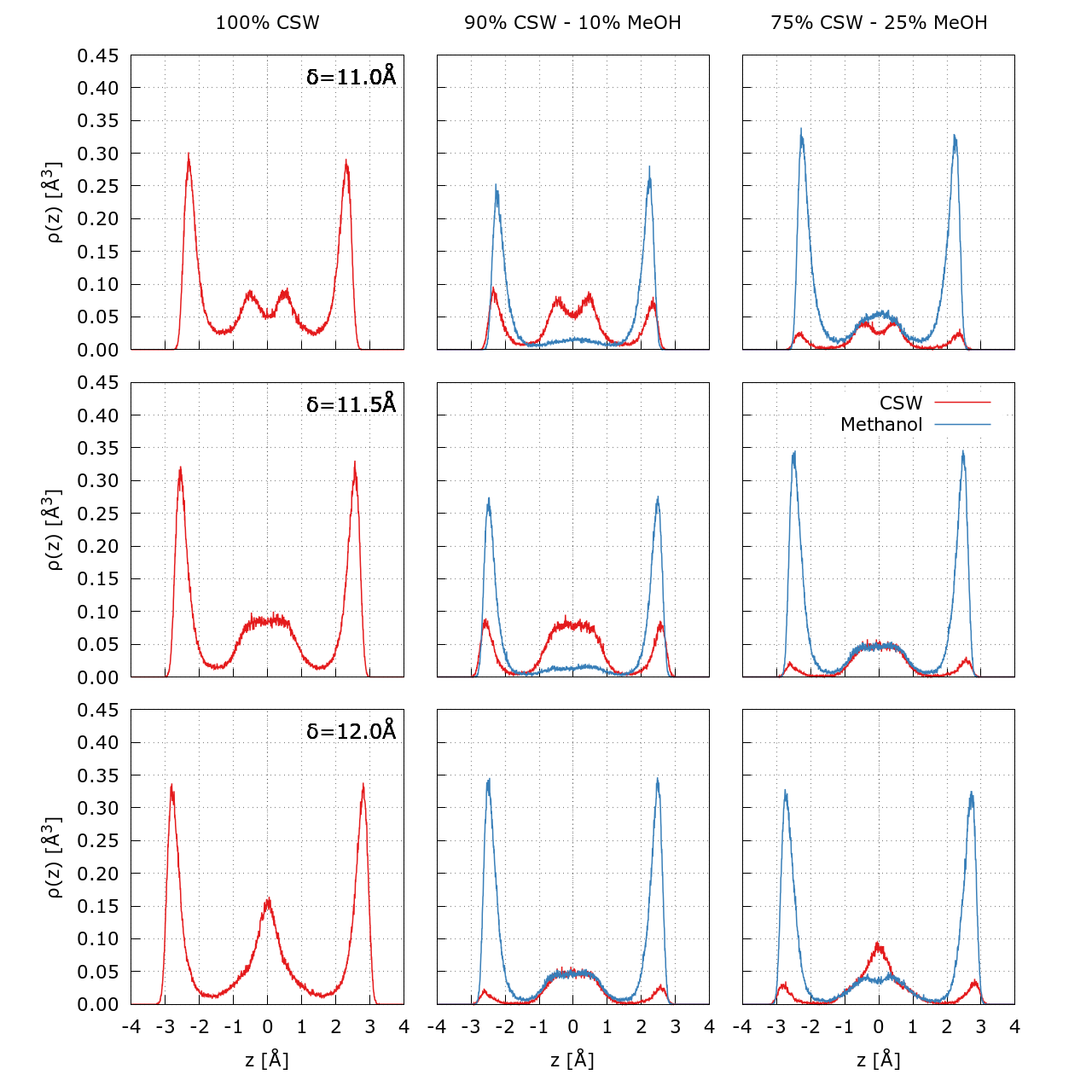}
\caption{{\bf Density profiles in the slit pore with a third CSW layer in the center.}
As in Fig.~\ref{fig:S2}, but for 
$\delta =11.0$\AA\ (upper panels), 
11.5\AA\ (central panels), and 
12.0\AA\ (lower panels).} 
\label{fig:S5}
\end{figure}

\begin{figure} 
\centering
\includegraphics[width=1\columnwidth]{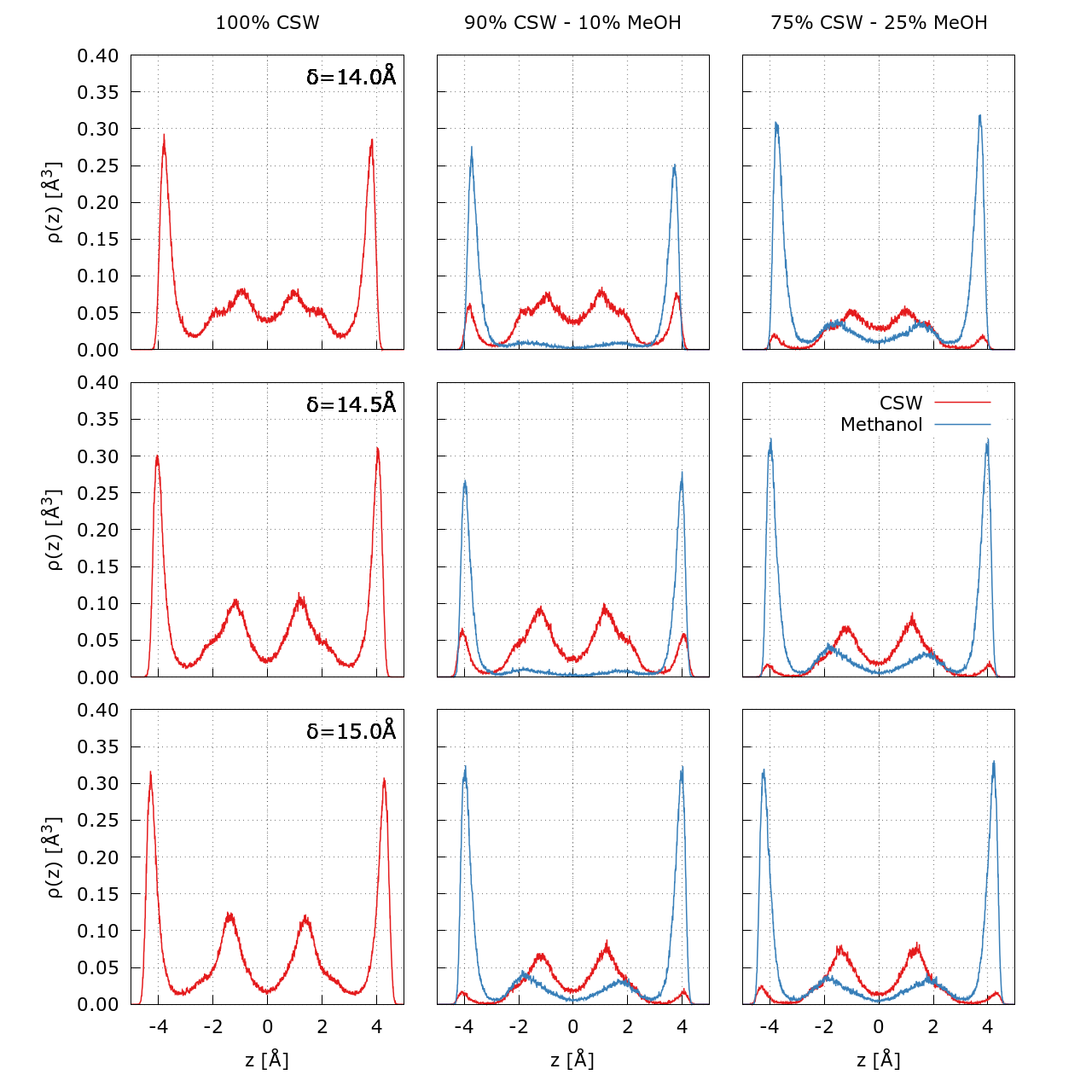}
\caption{{\bf Density profiles in the slit pore with a third CSW layer in the center.}
As in Fig.~\ref{fig:S2}, but for 
$\delta =14.0$\AA\ (upper panels), 
14.5\AA\ (central panels), and 
15.0\AA\ (lower panels).} 
\label{fig:S6}
\end{figure}

\begin{figure} 
\centering
\includegraphics[width=1\columnwidth]{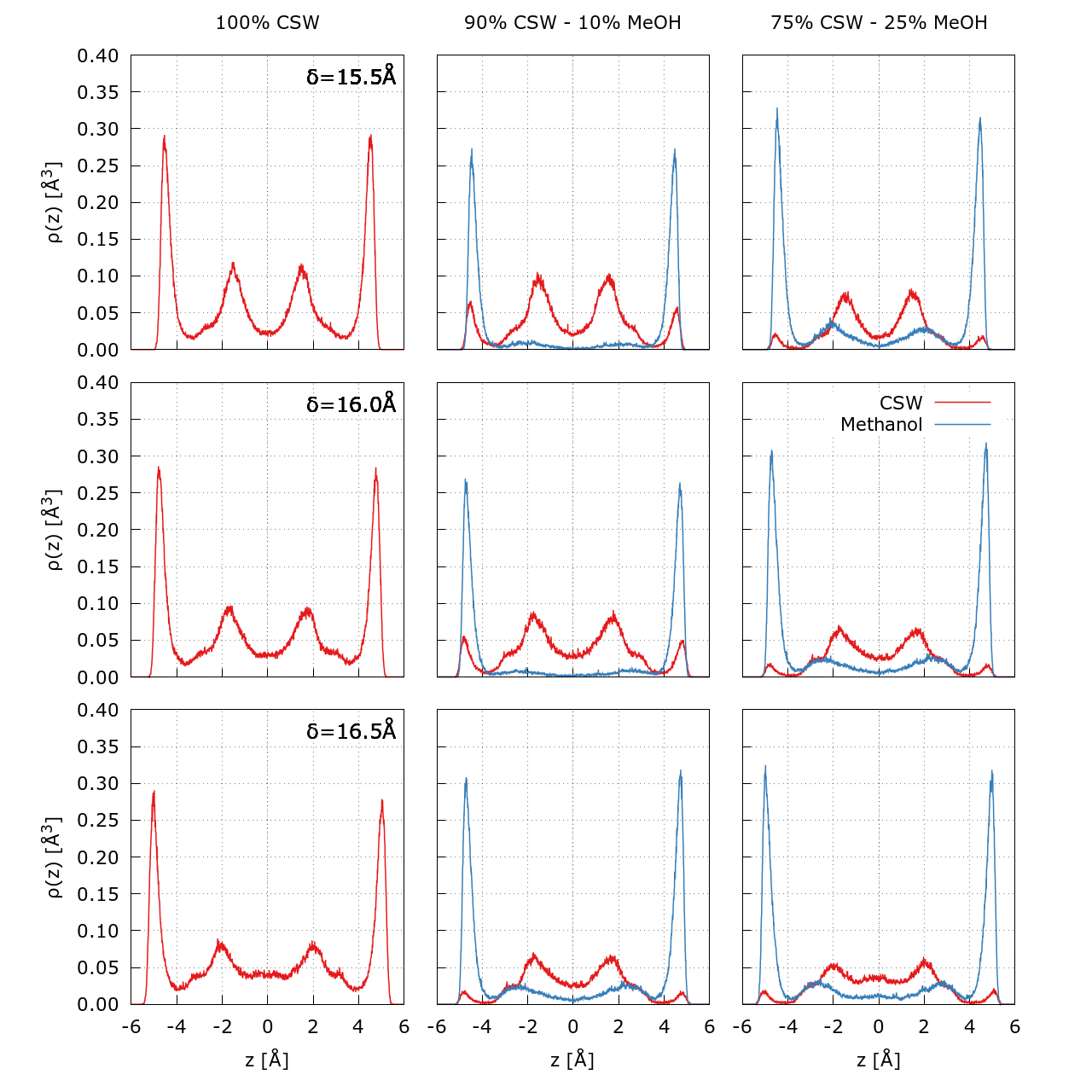}
\caption{{\bf Density profiles in the slit pore with a third CSW layer in the center.}
As in Fig.~\ref{fig:S2}, but for 
$\delta =15.5$\AA\ (upper panels), 
16.0\AA\ (central panels), and 
16.5\AA\ (lower panels).} 
\label{fig:S7}
\end{figure}

\begin{figure} 
\centering
\includegraphics[width=1\columnwidth]{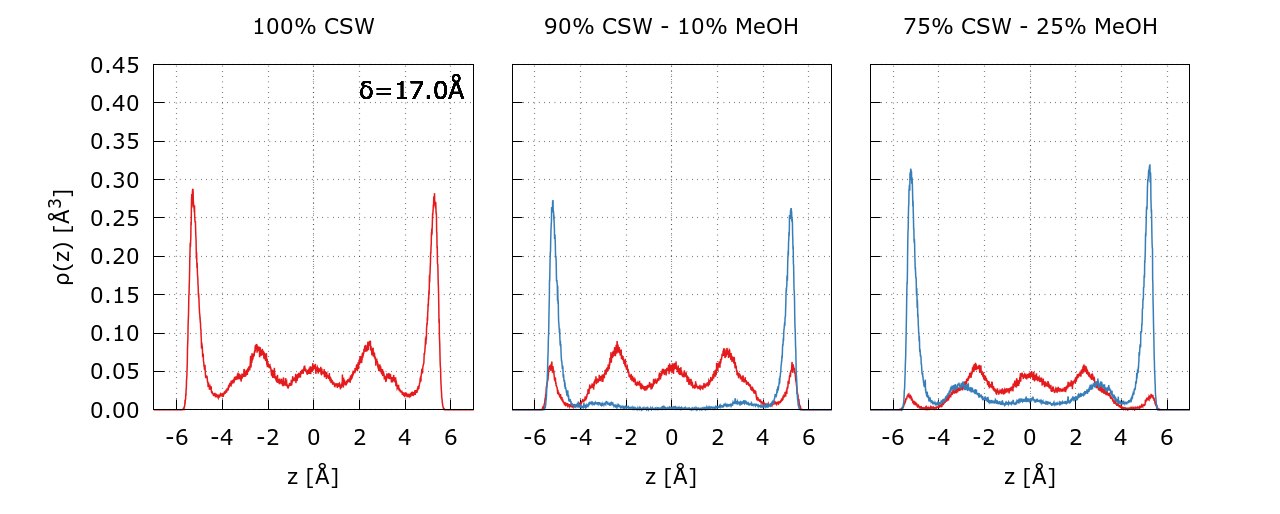}
\caption{{\bf Density profiles in the slit pore with a third CSW layer in the center.}
As in Fig.~\ref{fig:S2}, but for 
$\delta =17.0$\AA.} 
\label{fig:S8}
\end{figure}

\end{document}